\documentclass[review]{elsarticle}
\usepackage{tabularx}
\usepackage{paralist}
\usepackage{xcolor}
\usepackage{amssymb,amsmath,amsthm}
\theoremstyle{definition}
\newtheorem{defn}{Definition} 

\usepackage{multirow}

\usepackage{pdflscape,array,booktabs}
\usepackage{rotating}
\usepackage{epstopdf}
\usepackage[inline, shortlabels]{enumitem}
\usepackage{lineno,hyperref}
\modulolinenumbers[5]

\journal{Computer Science Review}

\begin{document}

\begin{frontmatter}

\title{Big Networks: A Survey}

\author[addr1]{Hayat Dino Bedru}
\author[addr1]{Shuo Yu}
\author[addr1]{Xinru Xiao}
\author[addr2]{Da Zhang}
\author[addr1]{Liangtian Wan}
\author[addr1]{\\He Guo}
\author[addr1,addr3]{Feng Xia\corref{mycorrespondingauthor}}
\cortext[mycorrespondingauthor]{Corresponding author}
\ead{f.xia@ieee.org}

\address[addr1]{Key Laboratory for Ubiquitous Network and Service Software of Liaoning Province, School of Software, Dalian University of Technology, Dalian 116620, China.}
\address[addr2]{Department of Electrical and Computer Engineering, University of Miami, USA}
\address[addr3]{School of Science, Engineering and Information Technology, Federation University Australia, Australia}

\begin{abstract}
A network is a typical expressive form of representing complex systems in terms of vertices and links, in which the pattern of interactions amongst components of the network is intricate. The network can be static that does not change over time or dynamic that evolves through time. The complication of network analysis is different under the new circumstance of network size explosive increasing. In this paper, we introduce a new network science concept called big network. Big networks are generally in large-scale with a complicated and higher-order inner structure. This paper proposes a guideline framework that gives an insight into the major topics in the area of network science from the viewpoint of a big network. We first introduce the structural characteristics of big networks from three levels, which are micro-level, meso-level, and macro-level. We then discuss some state-of-the-art advanced topics of big network analysis. Big network models and related approaches, including ranking methods, partition approaches, as well as network embedding algorithms are systematically introduced. Some typical applications in big networks are then reviewed, such as community detection, link prediction, recommendation, etc. Moreover, we also pinpoint some critical open issues that need to be investigated further.
\end{abstract}


\end{frontmatter}


\section{Introduction}

Complex systems are extraordinarily important in the current and near future \cite{tsiotas2019network}. Researchers of various fields consider the formulation of complex systems as a crucial issue. Complex systems are sometimes described by networks that are represented by nodes (vertices) and edges (links). Generally, nodes represent the entities and edges represent the connections amongst entities in the network, respectively. There are some examples of complex networks such as brain structures, transportation, mobile communication, social relationship, protein-protein interaction, etc. It has been proved that there exist different types of structural models, including scale-free, random, small-world, and regular networks \cite{garrido2011survey}.

There are numerous studied that investigated fundamental concepts in complex networks. Yu \emph{et al.}~\cite{yu2016networking} presented an in-depth survey of big data and technologies that are considered to be fundamental in big data. Specifically, they have elaborated the definition of big data, how to establish and illustrate big data as well as its available applications, including system modeling and big data scheduling. In this survey, the authors mainly focused on the hardware networking structure of big data. Xia \emph{et al.}~\cite{xia2017big} comprehensively surveyed big scholarly data, including its background and state-of-the-art technologies. They have discussed big scholarly data management as well as data analysis mechanisms, including social network analysis, content analysis, and statistical analysis. Besides, they have explained several big data technologies, such as academic recommendation systems and academic impact evaluation techniques. Similarly, Khan \emph{et al.}~\cite{khan2017survey} investigated the trends and challenges of big data from the perspectives of data management, analysis as well as data visualization. Additionally, Kong \emph{et al.}~\cite{kong2019academic} provided in-depth explanation of academic social network (ASN). They have discussed the background and relevant technologies of ASN. Furthermore, they presented detailed explanation of tools and models which can be suitable for ASN. These survey papers \cite{xia2017big,khan2017survey,kong2019academic} mainly focused on academic related data (e.g., DBLP\footnote{https://dblp.org/} and MAG\footnote{http://research.microsoft.com/en-us/projects/mag/}).

In this paper, we propose the concept of big networks (BNs) that are both complex and large-scale networks with higher-order and complicated inner structures. Analyzing the structure as well as characteristics of big networks is the most promising research issue in the area of network science \cite{steinbock2019analytical}. Furthermore, it is fundamental to understand the network topology in order to discover the classes and nature (i.e., static or dynamic) of a big network. However, how to characterize the structural form of BNs is an issue that needs serious attention from scholars. We analyze the structural characteristics of BNs from three levels including micro-level, meso-level, and macro-level. Also, the high-order algorithms are considered to find out the problems in BNs. Hence, we are motivated to propose a guideline framework that characterizes the main research areas of BNs.

Existing methods and algorithms have not specified BN issues in detail. Hence, this study aims to give guidance to researchers of big networks domain as well as providing insights into the basis of network science objects, from nodes to motifs. Therefore, we introduce basic ideas and explanations of big networks, review the up-to-date of network motifs detection algorithms, multi-layer networks, community detection, link prediction, recommendation methods, as well as the challenges occurred in these topics and open issues from the viewpoint of BNs.

This paper is structured as follows. Section \ref{structures} discusses the BN structure characterize levels. Section \ref{bnmodels} and \ref{techBNs} present the big network models and technologies in BNs, respectively. Section \ref{BNapps} introduces some of the important applications in BNs. Following the open issues and challenges of BNs in Section \ref{openchallenges}, we conclude the paper in Section \ref{conclusion}. 

\section{Structural Characteristics} \label{structures}

Researchers try to understand how communities/groups of individuals are densely connected with each other. Network models tend to focus on the network structures, and nodes inside the network are considered as individuals. At some point, it focuses on discovering the pattern of groups' connection. On the other hand, as social networks (SNs) become complex, a comprehensible pattern emerges from the local relationships of the network.

Social network analysis tends to focus on the scale relevant to the theoretical research area of the scholars. For instance, in a co-authorship network, one could analyze how weak or strong is the collaboration tie of individual authors, how big is a certain team or community in a network, and how concentrated is the tie strength~\cite{petersen2015quantifying}. There are three approaches to investigate and understand the network structure and characteristics: micro-, meso-, and macro-levels. These analysis levels are predominantly used in social science studies like sociology, political science, and economics.

At the \textbf{micro-level}, researchers analyze the node- and edge-levels of connection. In essence, it tends to focus on individuals and their associations to others. For instance, in a co-authorship network, analyses of micro-level might include a one-to-one link between authors. At the \textbf{meso-level}, researchers investigate group-level interactions that might include the characteristics of the group and how it is organized. Contrarily, at the \textbf{macro-level}, the analyses cover global characteristics of a given network. For instance, investigating the scientific collaboration of two different institutes found in geographically dispersed locations considered to be a macro-level. Moreover, scholars working in different levels investigate several features of scientific teams, propose distinct findings, and make contributions in terms of presenting numerous techniques and theories. Consequently, each level analyzes the different scale of data; adopt various methods, algorithms, and visualization tools.

\subsection{Micro-level}

At the Micro-level, we take into account individuals or a small group of individuals' interactions. For instance, the dyadic level considers communications among two people. Node-centric interaction is among the smallest unit of social network analysis. Moreover, micro-level examines the characteristics of individuals in a network. It also assesses the smallest levels of interaction between couples of vertices. It may also analyze the perception of how a certain vertex influenced by its connections.

\subsubsection{Vertices}

In mathematics, the network is a graph or a family of graphs that includes vertices and the set of interconnections between vertices. Usually, a set of vertices in a network $G$ represented as $V$ or $V(G)$. The vertices could be people in a social network, proteins in a biology network, and web pages on the internet. In single-layer networks, various measurements (such as PageRank, degree, closeness, betweenness, and eigenvector centrality) can be applied to identify influential nodes and analyzing the structural significance of each node \cite{gomez2019centrality}. When the characteristics are extended to multi-layer networks, they become different. For example, the degree of a node becomes a vector.

\subsubsection{Edges}

An edge is an interconnection that appears between two nodes, which can be weighted or unweighted and directed or undirected. A set of edges in a network $G$ commonly illustrated as $E$ or $E(G)$. Edges can build a complex structure in networks. The edges in the network model can be divided into three categories \cite{brug2018network}. \begin{enumerate*} [(1)] \item Explicit edges: These edges are known in networks, such as the ``following'' relationships in Facebook and ``referring'' relationships in citation networks. \item Discrete edges: These edges represent transactions between two nodes, such as text messages and phone calls. \item Inferred edges: These edges denote some statistical measure of similarity \end{enumerate*}. Since the data in the real world are often rich but noisy and sometimes even missing information, researchers gradually paid more attention to non-explicit edges. For instance, Newman \cite{newman2018network} proposed a technique that enables to provide optimal estimates of the accurate network structure by using rich but noisy data.

\subsection{Meso-level}

Meso-Level network analysis helps to understand better the nature of subnetworks, such as how subnetworks are formed, interactions between subnetworks, the difference between subnetworks, for instance, the number of vertices each subnetwork has and their features, and so on. Generally, it is a study of communities in the same network. It may also consider exploring networks that are particularly constructed to divulge links between micro- and macro-levels. Furthermore, meso-level networks might manifest the connection processes different from micro-level networks.

\subsubsection{Motif}

Network motifs are frequently recurring sub-graphs in a network whose distribution can reflect structural properties of complex networks \cite{simberloff2019network}. Because a motif can be regarded as a basic building block in the global system, it has important applications in many fields. For example, in \cite{muki2017algorithm}, the researcher applied it to the algorithm of constructing directed and unweighted networks. The algorithm starts from the empty graph and continues to select the in-degree or out-degree distribution of the network by encouraging or suppressing the formation of specific motifs. Besides, the discovery of motifs has also been applied in many fields, such as the functional analysis of brain neural networks in brain science, the pattern detection in biological networks, and the community discovery in social networks \cite{li2017motif,hu2019discovering}. As a result, motif discovery algorithms have gradually become active research topics in data mining.

There are two main types of existing motif discovery algorithms \cite{lin2017network}. \begin {enumerate*} [(1)] \item Based on Subgraph Enumeration: Algorithms under this category are not effective in finding motifs with more than eight nodes \cite{sun2019efficient}. \item Based on Frequency Estimation: Compared with the first type, algorithms which lie under this category can get a better result in finding large motifs. However, they generally cost too much computing resources \cite{al2019online}. To deal with this problem, Lin \emph{et al.}~\cite{lin2017network} proposed a solution based on GPUs (Graphical Processing Units) to reduce the overall computational time, which parallelizes a great number of tasks of subgraphs matching when calculating the frequency of subgraphs in random graphs. In the meantime, they also experimented on various biological networks; and obtained several key factors affecting GPU performance.\end {enumerate*}

\subsubsection{Hyper-edge}

Compared with the edges in a general graph, which can only indicate the connection between a pair of vertices, the hyper-edge in the hyper-graph can contain multiple vertices. Mathematically, a hyper-graph is a graph that can be used to represent the connection between multiple vertices. In a hyper-graph, an edge can be linked to any number of nodes, that is called hyper-edge. For instance, in a general network of scientific collaboration, the edge can only represent whether two authors have collaboration relationship. However, in a scientific collaboration hyper-network (network with hyper-graph topology), a hyper-edge can represent an article written by several authors.

Since the relationships in the real world are often not just simple binary relationships, the research studies on hyper-graph have gradually become a hot spot. The introduction of hyper-edge can reduce not only the complexity of the network structure but also portray more complex relationships. At present, many types of research on hyper-edges and hyper-graphs focused on the characteristics of hyper-network. For example, in \cite{purkait2017clustering}, the team of Purkait has proved theoretically and experimentally that using large hyper-edges can get better clustering accuracy in hyper-graph clustering, and has also proposed a sampling large hyper-edges algorithm. In \cite{kabiljo2017social}, Kabiljo \emph{et al.} proposed a distributed algorithm which can partition hyper-graph with billions of vertices and hyper-edges in a few hours.

\subsection{Macro-level}

Rather than individuals and communities interactions, in the macro-level we analyze the structure of large-scale as well as complex networks at the level of a network component, density, and so on. This is a deeper level that studies at the level of the whole big network.

\subsubsection{Network Density}

Network density assesses the density of edges between nodes in a network. It is also the quantitative relation of the total edges in the network to the maximum variable that the network can accommodate. It also explains the percentage of actual links that could appear between two vertices. The actual links are connections that exist in the network. For instance, in a particular scientific team, the actual links between researchers might be many (--- it might even be a 100\% of all possible links in the team). A possible link is a link between researchers that might exist in the network. On the other hand, the actual link between researchers is likely to be low in comparison to possible links that appear at a conference. Hence, we could say that the network density in a scientific team is high but relatively low density at the conference.

Network density $D$ for an undirected network is mathematically represented as $D=\frac{2E}{N(N-1)}$, where $N$ and $E$ refer to the number of nodes and edges in the network, respectively. As an essential parameter in network science, it is mainly applied as an evaluation criterion in experiments \cite{brug2018network}.

\subsubsection{Overlap and Multi-degree}

Since there are often overlapping links in networks, it is an important task to study the overlap and multi-degree. The overlap in the multi-layer network can be divided into two types: global overlap and local overlap \cite{bianconi2013statistical}. Global overlap between layer $\alpha$ and layer $\beta$ can be defined as: $O^{\alpha\beta}=\sum_{i<j}a_{ij}^{\alpha}a_{ij}^{\beta}$, where $\alpha \neq \beta$ and $a_{ij}^{\alpha}=\{1,0\}$. If $a_{ij}^{\alpha}=1$, it indicates the presence of a link between node $i$ and node $j$  in layer $\alpha$. Correspondingly, local overlap can be defined as: $o_{i}^{\alpha\beta}=\sum_{j=1}^{N}a_{ij}^{\alpha}a_{ij}^{\beta}$.

Multi-link for nodes in a multi-network defined as $\vec{m}=(m_{1},.., m_{\alpha},..,m_{M})$, where $m_{\alpha}=\{1,0\}$ and $m_{\alpha}=1$ represents that the nodes are connected in layer $\alpha$. Furthermore, we can infer that $\vec{m}=0$ if and only if two nodes are not linked in all layers. Therefore, we can define multi-adjacency matrix as  $A^{\vec{m}}$:$A^{\vec{m}}=\prod_{\alpha=1}^{M}[a_{ij}^{\alpha}m_{\alpha}+(1-a_{ij}^{\alpha})(1-m_{\alpha})]$, where $A^{\vec{m}}=1$ if and only if there is a multi-link between node $i$ and $j$. Thus, define multi-degree $\vec{m}$ of a node $i$, $k_{i}^{\vec{m}}$ as the total number of multi-links $\vec{m}$ connected to node $i$, that is, $k_{i}^{\vec{m}}=\sum_{j=1}^{N}A_{ij}^{\vec{m}}$ \cite{cellai2013percolation}.

\begin{figure}[htbp]
\centering
\includegraphics[width=0.5\textwidth]{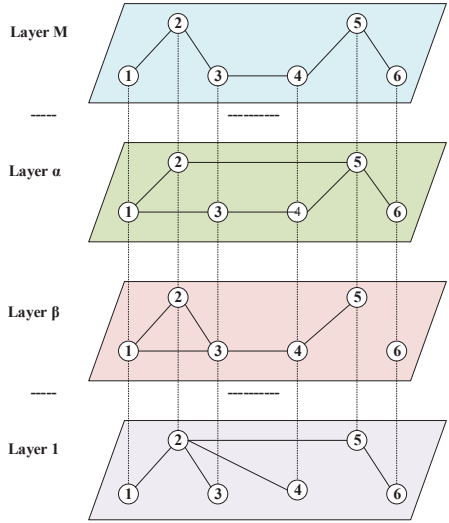}
\caption{The structure of a multi-layer network. }
\label{fig:mlayer}
\end{figure}

Suppose a network has four layers, as shown in Fig. \ref{fig:mlayer}, and assume the layers from bottom to top labelled as 1-4. Since there are edges between node 1 and node 2 in four layers, the multi-link of node 1 and node 2 is (1, 1, 1, 1). Likewise, there are edges between node 2 and node 5 in layer 1 and layer 3. However, in layer 2 and layer 4, there are no edges between them. Thus, the multi-link for them is (1, 0, 1, 0).

Besides the overlap of links, there might exist an overlap of motifs as well as an overlap of communities in a network. Li \emph{et al.}~\cite{li2017motif} combined the motif discovery technique and clustering to discover overlapping communities in social networks and achieved good experimental results.

\section{Big Network Models} \label{bnmodels}

In this section, we give comprehensive reviews of various big network models, including time-aware BN model, motif-based BN model, and multi-layer BN model. In each subsection, we discuss the overview of each model, categories of the models, and their corresponding algorithms from the perspective of BNs.

\subsection{Time-aware Big Network Model}

A network is a prevalent form of representing information. For instance, in a social network, there is a form of graph that is connecting people, in biological networks, there are regulatory structures, influences, and correlations in the form of a graph, and in academic social networks, there are researchers linked through citations or co-authorship~\cite{kong2019academic}. Networks can be static, where the vertices and links do not change over time, or dynamic, where both can appear or disappear throughout the lifetime of the network.

Furthermore, in a static network, there is no change in vertices, and links remain the same permanently. Whereas in a dynamic network, there is a probability of vertex disappearance and the formation of new vertex. The disappearance may occur in their links although they can be recovered or reappeared. Also, the topological structure of dynamic networks varies over time. Some examples of real-world dynamic networks are social networks, transportation networks, and communications networks.

In this section, we present summaries of static and dynamic networks. We focus on the high-level topics that are crucial in big networks. For more comprehensive reviews, readers can refer to \cite{li2017fundamental,masuda2016guidance,farine2018choose,michail2018elements}.

\subsubsection{Static Network Model}

\begin{figure}[!h]
  \centering
  \includegraphics[width=7cm]{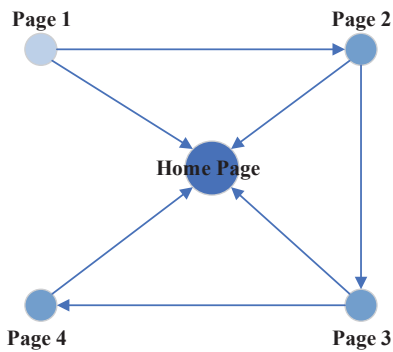}
  \caption{A static web network that is directed. The connection between vertices (i.e., web pages) depict the hyperlinks}\label{StaticNet}
\end{figure}

The contents in a static network either rarely or never changes. For instance, if we take a static website, the contents on it remain there for days, weeks, months, or even for years (see Figure \ref{StaticNet}). The nature of a static network can be undirected or directed and unweighted or weighted.

As stated in \cite{masuda2016guidance}, there are two fundamental ways to represent static networks; these are adjacency matrix as well as link list. These representations highlight features of static networks, and are susceptible to specific kinds of computations. In the adjacency matrix, networks can be illustrated as an $N\times N$ matrix, in which two vertices are adjacent if they have links between them that connect them directly. Note that, representing a static network using adjacency matrix is beneficial while developing and quantifying the structure and dynamical processes of the network. However, it consumes much memory at the time of computation. The processing of a network with $N$ number of vertices requires a complexity of \textsc{O}($N$). Having considered the limitation, the link list can be an option to represent a static network. Unlike the adjacency matrix, the link list is efficient to use for randomization of links as well as for numerical experiments of networks with sparse interactions.

There are numerous mechanisms utilized to analyze the structure and characteristics of a static network starting by measuring some of the properties of networks. For instance, \begin{enumerate*} [(i)] \item analyzing degree distribution to describe the connectivities between networks, \item the average path length in the network so that one can tell how fast information can propagate, and \item clustering coefficient to find out the group fitness of individuals in the network \end{enumerate*}. Quantifying such statistics is a non-trivial task; hence, there are more sophisticated methods to analyze networks. In some cases, data analyst are interested in analyzing something called local network property, which is calculating the frequency of occurrences of subgraphs in a network, i.e., network motifs (see Section \ref{motifdesc}). Similarly, to evaluate the importance of vertices in a network, analysts employ several measurements such as PageRank, Katz, degree centrality, betweenness centrality, as well as closeness centrality \cite{gomez2019centrality}.

Furthermore, one of the most crucial issues in big network analysis is analyzing the community structure of a network \cite{fortunato2016community}. Thus, scholars proposed numerous approaches to discover communities in a static network; one of the well-known methods is Infomap. Infomap is designed explicitly for a directed and weighted static network that aims to identify the non-overlapping community structure of a network. There are also methods that detect the overlapping communities of static networks such as \emph{K}-clique algorithm and the Lancichinetti method \cite{masuda2016guidance}.

Rand \emph{et al.}~\cite{rand2014static} studied the usefulness of static network in the context of human cooperation. The authors claimed that a static network structure helps to make human cooperation steadfast. Verily in a fixed type of network, interactions among cooperators become more intense in such a way that they benefit each other more. Rand \emph{et al.}~\cite{rand2014static} presented evidence that supports the argument that static networks can promote human cooperation.

\subsubsection{Dynamic Network Model}

\begin{figure}[h!]
  \centering
  \includegraphics[width=7cm]{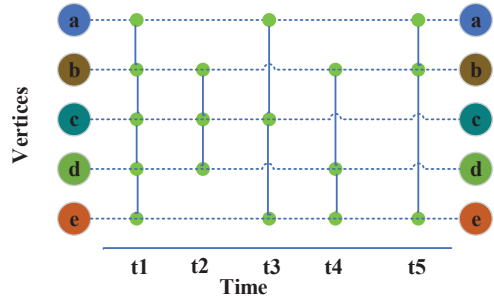}
  \caption{Dynamic Network with 5 number of vertices showing the evolvement of interactions among vertices in different time spans}\label{DynamicNet}
\end{figure}

Networks that evolve over time are called temporal or dynamic networks, such as transportation networks, social networks, communication networks, networks of citations, and many more real-world networks \cite{michail2018elements,boccaletti2006complex}. As stated in \cite{rubenstein2015similar}, in dynamic networks, connections are denoted by a time-slot of static networks. In essence, in contrast to static networks, dynamic networks consider the timestamps as well as take into account the temporal information. Figure \ref{DynamicNet} shows a simple example of a dynamic network.

From the perspectives of human behaviors, Rand \emph{et al.}~\cite{rand2011dynamic} discussed that in dynamic networks, changes occur regarding the behavior of an individual's connections in a social network. Moreover, the authors found out that human cooperation decreases through times when the random-walking process takes place in social networks. Additionally, human cooperation will decrease or increase when there are infrequent and frequent changes in the network, respectively. However, the experimental results in \cite{rand2011dynamic} indicate that the dynamic nature of social networks can promote human cooperation in large groups of interactions.  Similarly, Melamed \emph{et al.}~\cite{melamed2018cooperation} proved that dynamic networks endorse cooperation at the higher levels where there is a new formation of connections or else discarding of a connection.

Analyzing the structural characteristics of a dynamic network as well as measuring its properties has the same purpose and features as of a static network. However, researchers extended the models and methods proposed for static networks so that they could fit in dynamic networks. For instance, Luis \emph{et al.}~\cite{rocha2014random} proposed a random-based measurement to quantify the centrality of individuals in a temporal network called TempoRank. TempoRank is an extension of PageRank that mainly works for static networks. In \cite{holme2013temporal}, the authors categorized the centrality measures of vertices for dynamic networks into two, such as time-dependent and time-independent centrality measures. The former identifies the changes in the importance of a vertex. Also, it analyzes the probability that a vertex influential at a particular time may not be influential at other times. Whereas the latter evaluates how a vertex is vital in general. Recently, Koo \emph{et al.}~\cite{koo2019incremental} proposed a ranking algorithm specifically for a dynamic web environment.

Like static networks, one of the challenging tasks in a dynamic network is community detection. Moreover, it is vital to analyze the structure of the interactions of vertices and how they evlove at times. Hence, Liu \emph{et al.}~\cite{liu2018global} proposed a community detection method for dynamic networks called ``persistent communities by eigenvector smoothing (PisCES)'' which is derived from degree correction (\textemdash{heterogeneity of degree within clusters}) and evolutionary spectral clustering techniques. The method merges information across a sequence of networks over time. In another work \cite{sarmento2019dyncomm}, scholars proposed an R package dynamic community detection for evolving networks called DynComm. DynComm has an understandable application programming interface (API) that eases the detection of communities for a big dynamic network\cite{aggarwal2014evolutionary}. Table \ref{tab:StatDyn} briefly shows the comparison of static and dynamic networks.
\begin{table*}[ht]
  \centering
  \caption{Comparison of Static and Dynamic Networks}
\begin{tabular}{p{9em}lp{13em}}
\cline{1-3}\multicolumn{1}{r}{} & Static Network & \multicolumn{1}{l}{Dynamic Network} \\
\hline
\multicolumn{1}{l}{Overview} & \multicolumn{1}{p{10em}}{information either rarely or never changes } & information/data evolve and change over time, important to disclose patterns that might be hidden in a more aggregated network \\
Centrality Measurements & \multicolumn{1}{p{10em}}{Degree Centrality, PageRank, Katz, and other classic measurements \cite{gomez2019centrality}} & \multicolumn{1}{l}{TempoRank~\cite{rocha2014random}, \newline{}C-Rank~\cite{koo2019incremental}} \\
Community Detection Methods & \multicolumn{1}{p{10em}}{Infomap, Fast Unfolding Method \cite{masuda2016guidance} } & \multicolumn{1}{l}{DYNCOMM \cite{sarmento2019dyncomm}, \cite{abrahao2014separability}, \cite{cordeiro2016dynamic}} \\
Overlapping Community Detection Methods & \emph{K}-Clique \cite{masuda2016guidance}  & \cite{marquez2019overlapping}, \cite{sariyuce2016sonic} \\
\hline
\end{tabular}%
 \label{tab:StatDyn}%
\end{table*}

\subsection{Motif-based Network Model} \label{motifdesc}

Recently, network motifs are getting more attention from researchers as network motifs are useful to discover the structure of big networks \cite{masoudi2012building}. Researchers are adapting the concept of network motifs to analyze the structure of big networks including social networks, co-authorship networks, biological networks, neural networks, protein-protein interaction networks, and so on. A variety of networks inclined to have various collections of local structures that occur frequently \cite{simberloff2019network}. In this section, we discuss network motifs, specifically the concept of network motif and the algorithms of discovering network motifs in different scenarios within big networks.

The theoretical definition of network motif is first proposed by Milo \emph{et.al} \cite{milo2002network}, wherein, they described network motifs as ``patterns of interactions occurring in complex networks at numbers that are significantly higher than those in randomized networks''. Generally speaking, if the frequent occurrence of a subgraph $g'$ in a network $G$ is more than it occurs in a random network, then $g'$ will be labeled as a network motif.

Network motifs help to understand big networks by identifying small functional subgraphs. Those subgraphs are simpler to understand in contrast to the whole complexity of the big network at once. The subgraphs described by certain patterns of interactions among nodes may show efficiently achieved structural characteristics of a particular network.

Milo \emph{et.al} \cite{milo2002network} discussed network motifs in a food web network assuming a directed uni-partite network in which vertices and links represent the group of species and the flow of energy through the network, respectively. Moreover, it essentially looks for common patterns that are occurring between three species. Furthermore, having considered the limitation of studies regarding network motifs in dynamic networks, Paranjape \emph{et al.}~\cite{paranjape2017motifs} introduced a notion that gives insights into the importance of motifs in networks that evolve over time. They explained temporal motifs as ``induced subgraphs on a sequence of temporal edges''. Also, they proposed an algorithm that counts available motifs in a given temporal network.

Researchers have proposed several algorithms to identify patterns of reoccurring interactions and essentially see which ones occur more frequently than expected randomly. In this paper, we discuss two types of motifs, including triangle motifs and higher-order motifs. Besides, we present existing algorithms that tackle network motifs discovery challenges by taking into account the complexity and size of the networks. Moreover, the algorithms discussed here are selected approaches that can be comparatively applicable to BNs.

\subsubsection{Triangle Motif}

Triangle Motifs could appear in a particular network that designates the interactions among three vertices. Moreover, it is beneficial to comprehend the inter-connectivity of vertices in a network. Also, a triangle motif describes the social pattern in a network \cite{stone2019network}. It can also model a social closure. Let us consider a static directed network $S$ that is induced by links of motifs $T$. In triangle motifs, $S$ comprises 3 vertices and at least one directed link amongst any pair of vertices. $S$ of $T$ consists at least three and at most six static edges \cite{paranjape2017motifs}.

\subsubsection{Higher-order Motif}

The high order network structure is associated with a graph and subgraph. In complex networks, the number of motifs is calculated for graph clustering and community detection. The higher-order motifs are computed to find the relation in pair of the nodes and the authority of the nodes \cite{zhao2018ranking}. High order connectivity pattern are building blocks of a single homogeneous network which are essential for the modeling components of the network. A graphlet is a small connected subgraph, and the non-trivial graphlet is a node pair structure connected by an edge. Higher order graphlets have a greater number of nodes and edges.

\begin{figure}[h!]
  \centering
  \includegraphics[width=9.5cm]{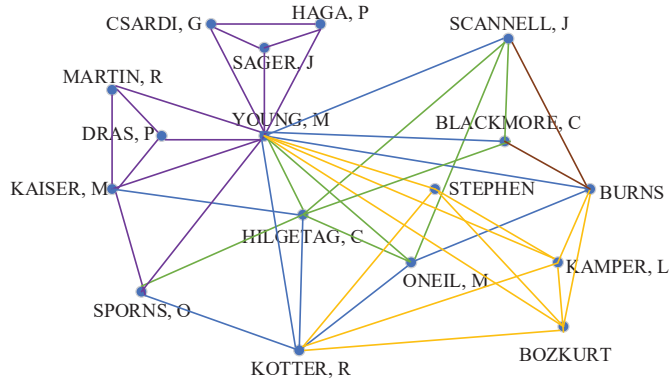}
  \caption{The high-order network visualization of a small co-authorship network. The different colors of the edges represent different high-order motifs that appear in the network}\label{fig:highorder}
\end{figure}

Further, a typed network is used to uncover the high order organization of heterogeneous networks. The typed graphlet network captures both the connectivity pattern and typed \cite{ritchie2014higher}. An imperative high order network structure such as cliques and big stars can be discovered interactively by the user in real-time. Network motifs noticeably identify the vital higher-order structures. Figure \ref{fig:highorder} shows the higher-order network structure of a small co-authorship network.

\subsubsection{Motif Discovery Algorithms}

The baseline motif discovery approaches presented at the early stage primarily consider two fundamental stages:  \begin {enumerate*} [1)] \item calculating the frequency of all subgraphs of a certain amount obtained in the network known as ``subgraph census''; \item generating a set of similar random graphs with similar degree sequence like the given network.\end {enumerate*} At the second stage, the subgraph census is computed on each of generated subgraphs from which the statistical significance of isomorphic subgraphs of distinct classes is computed as well. The statistical significance is computed by using the probability of patterns being overrepresented. The main limitation of such methods occurs while computing subgraphs census even in a network with less number of nodes. Thus, in this section, we discuss recently proposed algorithms that take into account the limitation mentioned above as well as computational complexity while applied in a big network.

\subsubsection*{gLabTries}

G-tries is a prefix tree data structure that facilitates the storage of a set of graphs efficiently by preventing re-use of the subgraphs information among common prefixes. Misael \emph{et al.}~\cite{mongiovi2018glabtrie} proposed motif discovering algorithms for both undirected and directed networks called gLabTrie. gLabTrie is an extension to the original G-tries motif discovery algorithm \cite{ribeiro2014g}. gLabTrie is a data structure for discovering motifs with constraints. As stated in \cite{mongiovi2018glabtrie}, the performance of this method highly depends on a certain network size. The fundamental change made on gLabTries is ``label-based queries''. Mongiov{\'\i} \emph{et al.}~\cite{mongiovi2018glabtrie} defined label-based queries as quadruple $Q$ containing multiset of labels $C$, requested size of motifs $k$, frequency threshold $f$, and $p$-value threshold $(Q = (C, k, f, p))$. While implementing gLabTrie, users give sets of constraints as a requirement, and the system generates topology for each specified constraints.

\subsubsection*{VALMOD: Variable Length MOtif Discovery}

To mine network motifs discovery of variable lengths, Linardi \emph{et al.}~\cite{linardi2018matrix} proposed an algorithm called VALMOD. This algorithm has the ability to discover the top-$k$ motifs pairs of variable length. VALMOD is a scalable algorithm that can be used by users to reveal accurate motifs efficiently. Besides the motif discovery algorithm, they also proposed motifs ranking approach named as VALMAP. VALMAP is a metadata series that mainly uses a new normalized length for ranking motif pairs of variable length.

\subsubsection*{LCNM: Large Co-regulatory Network Motif}

Luo \emph{et al.}~\cite{luo2018efficient} proposed an algorithm named large coreglulatory network motif (LCNM) that aims to detect large coregulatory motifs with relatively low computational complexity. They mainly considered colored network motifs in a large human coregulatory network. Moreover, Luo \emph{et al.} proposed candidate subgraphs patterns generating methods such as quick sampling and random walking methods as well as exhaustive counting to generate all subgraph patterns. The authors adopted G-tries aiming to make the algorithm capable of saving a set of motifs in G-tries. Moreover, G-tries is improved in such a way that it could identify the maximum number of motifs of a size larger than 4 nodes in a large network. Besides, a method that improves the computational complexity of motif discovery in a large network is also proposed~\cite{luo2018efficient}. However, it still consumes time when applied it to a big network with thousands and millions of nodes. Unlike other methods, LCNM can be able to discover motifs up to a maximum of 8 interacting nodes.

\subsection{Multi-layer Big Network Model}

Recently, multi-layer networks (MLNs) are getting attention from scholars in many disciplines, including economics, infrastructures, climate, neuroscience, and so on. MLNs have been presented under the circumferences of social sciences to explain distinct types of social interactions existing among the vertices of social networks~\cite{bianconi2018multilayer}. More than one interrelating networks form a multi-layer network, and one typical example of MLNs is a social network~\cite{dickison2016multilayer}. Describing MLNs is critical to comprehending complex and big networks such as brain networks \cite{de2017multilayer}, transportation networks \cite{cardillo2013modeling}, big scholarly networks, and so forth. Also, MLN makes it easier to characterize the structure of big networks. Furthermore, it provides a comprehensive perspective of big networks compared to the framework of a single layer network \cite{bianconi2018multilayer}.

\begin{defn} \textit{A multi-layer network has a set of vertices, edges, and layers $G (V, E, L)$. The layer is the one that contains different characteristics of a given network. Moreover, it is a combination of networks at different layers with distinct types of edges (i.e., multiple types of interactions) among vertices.} \end{defn}

Also, Bianconi~\cite{bianconi2018multilayer} defined a multi-layer network as follows.

\begin{defn}\textit{A given multi-layer formed by distinct $M$ layers is formed by a set of $M$ networks describing the interactions within each layer and $M(M-1)/2$ networks describing the interactions between nodes in every pair of different layers.}\end{defn}

Additionally, in \cite{bianconi2018multilayer}, MLN is mathematically defined as:

\begin{defn}\textit{MLN is given by the triple $G_M =\left (\acute{Y} ,\acute{G} ,\c{G}\right )$, where $\acute{Y}$ denotes the set of layers, such that $\acute{Y} =\{\alpha  \mid \alpha  , \in \{1 ,2 ,\ldots  ,M\}\}$ of the MLN, and $M$ denotes the total number of layers, i.e., the cardinality of $M = \mid \acute{Y} \mid $.}\end{defn}

The network $G_M$ has $n$ number of vertices in each layer, $V=\{1,2,3,\ldots,n\}$, and $M$ layers with different characteristics. Each layer contains a set of vertices. The vertices can create links within the layers (i.e., intra-layer links) as well as across the layers (i.e., inter-layer links). For example, assume there is a scholarly multi-layer network with two layers in which the first layer is a citation network, and the second one is a co-authorship network. In the citation network, vertices and edges represent papers and the citing papers, respectively. In the co-authorship network, authors are vertices, and they get connected if they co-authored one or more papers together. The interactions that appear among these two different networks form an authorship network, i.e., authors linked to the papers they wrote.

The framework of MLN reduces the challenges that happen while measuring the centrality of the vertices, detecting communities, discovering influential communities, predicting links, and recommending in a big network.

\subsubsection{Community Detection in MLNs}

Mucha \emph{et al.}~\cite{mucha2010community} proposed the first community detection algorithm considering a multi-slice network. A multi-slice network is one kind of multi-layer networks in which a combination of different networks tied over connections that link each vertex from a specific slice to another. The proposed algorithm allows the analysis of the network's community structure that changes over time, i.e., a temporal network. The type of network considered in their study has several scales and links with distinct characteristics. The authors implemented their algorithm on different real-world networks, and have obtained satisfying results.

Additionally, in \cite{lancichinetti2012consensus}, another approach has been introduced mainly to identify consensus clusterings in a multi-layer network. The method produces accurate and stable results deriving out of partitions provided by stochastic approaches. Moreover, while combining the method with other existing community detection algorithms, it enhances the accuracy and stability of the generated partitions. Also, the authors claimed that the method is suitable to characterize and keep track of the community structure of temporal networks. Lancichinetti \emph{et al.}~\cite{lancichinetti2012consensus} applied the method on large-scale citation networks and witnessed its capability to control the structure of multi-layer networks.

De \emph{et al.}~\cite{de2015identifying} proposed an algorithm that generates overlapping communities in a multi-layer network, i.e., the method identifies communities across layers that instigated from similar interaction.

Furthermore, Raul \emph{et al.}~\cite{mondragon2018multilink} introduced a method that discovers the rich structure of communities of multi-layer networks by connecting each multi-link with a community. The multi-links portray the associations presents amongst vertices of the multi-layer networks, and they are a combination of a distinct number of appropriate layers.

\subsubsection{Quantifying Centrality and Vertex Ranking in MLNs}

Quantifying the centrality as well as the ranking of vertices in a multi-layer network is as critical as it is in a single layer network. Thus, numerous approaches have been proposed by interested scholars. Many of the measurements proposed to identify the importance of vertices in single layer networks are extended to be applied to multi-layer networks. For instance, the PageRank method is extended to Multiplex PageRank \cite{abrahao2014separability}, which assesses the centrality of vertices of multi-layer networks. Mainly, Multiplex PageRank evaluates how the central vertex in one layer is influential on another layer. For example, suppose we have a co-authorship network containing a collection of scholars who work explicitly on big data. Scholar $A$ is the prestigious scholar with high centrality score in this network. Thus, if $A$ takes part in another scientific team that works explicitly on cloud computing, the centrality score of $A$ will might have an impact. Additionally, it influences the centrality of $A$ in the other co-authorship network with a different research area. Hence, according to the experiment done by \emph{Abrahao et al.}~\cite{abrahao2014separability}, a vertex's centrality in a particular layer might affect the centrality of the same vertex in another layer.

Additionally, considering the limitation of Multiplex Pagerank, Rahmede \emph{et al.}~\cite{rahmede2017centralities} introduced an algorithm that effectively ranks vertices as well as layers of the MLNs. The centrality and importance of vertices are dependent on each other. Moreover, the authors argued that a layer with more central vertices in it attains a more significant influence than layers with less central vertices. Luis \emph{et al.}~\cite{sola2013eigenvector} extended the standard eigenvector centrality measure to be suitable to MLNs. The method measures the importance of vertices in MLNs.

\section{Technologies in Big Networks} \label{techBNs}

In this section, we introduce state-of-the-art technologies of BNs, such as ranking approaches, partitioning algorithms, as well as overview of network embeddings and available techniques.

\subsection{Ranking Algorithms}

The main idea of ranking is mining information available in the cloud or in any storage area. The aim of ranking is to extract data which are appropriate for the purpose they are intended for. Some of the instances which clearly show ranking impacts are: how recognizable are human's merit and success \cite{spitz2014measuring,waltman2016review}, how to distinguish and prevent an infectious disease while happening without warnings \cite{iannelli2017effective}, how to assign funding for scientific research, and how to identify key authors in multi-authored papers.

Understanding the network representation of any input data is a critical part of ranking algorithms. Nowadays, the complex network has risen as one of the main promising approaches to analyze different categories of complex data like financial, information systems, and social \cite{barabasi2016network}. As a result, network representation helps to minimize the complexity of any system. It also enables users to comprehend the structure and dynamics of any complex system.

There are abundant surveys and literature reviews that cover ranking methods \cite{liao2017ranking}. In this review, we discuss algorithms designed particularly for ranking vertices, motifs, and communities. Table \ref{tab:RankMethods} shows a summary of the ranking methods. Note that network type refers to weighted or unweighted and directed or undirected.
\begin{sidewaystable*}
  \centering
  \caption{Summary of Vertex Ranking Methods. }
\begin{tabular}{lllll}
\hline
\multicolumn{1}{p{8.5em}}{Method name/ Reference \#} & \multicolumn{1}{c}{ClusterRank \cite{chen2013identifying}} & \multicolumn{1}{c}{E-Burt \cite{hu2018ranking}} & \multicolumn{1}{c}{BridgeRank \cite{salavati2018bridgerank}} & \multicolumn{1}{c}{Ref \cite{wei2018identifying}} \\
\hline
Objective & \multicolumn{1}{p{11.28em}}{quantifying the influence of a node by considering its direct neighbors as well as its clustering efficient} & \multicolumn{1}{p{9.89em}}{measuring how influential a node is in the network} & \multicolumn{1}{p{11.22em}}{identifying nodes that are capable of spreading information at the fastest} & \multicolumn{1}{p{14em}}{computing vertices considering both overlapping and non-overlapping community structure of a network } \\
\multicolumn{1}{p{8.5em}}{Adopted technique} & \multicolumn{1}{p{11.28em}}{clustering coefficient} & \multicolumn{1}{p{9.89em}}{structure holes and connectivity strength} & \multicolumn{1}{p{11.22em}}{closeness centrality } & \multicolumn{1}{p{14em}}{network representation learning} \\
Proposed & \multicolumn{1}{p{11.28em}}{nodes ranking index that uses local information} & \multicolumn{1}{p{9.89em}}{nodes ranking method} & \multicolumn{1}{p{11.22em}}{nodes ranking method that consider local closeness centrality value} & \multicolumn{1}{p{14em}}{vertex ranking method} \\
Network type & \multicolumn{1}{p{11.28em}}{Directed and can also work in an undirected} & Weighted & undirected and unweighted & works in all types \\
Ranking level & vertex & vertex & vertex & vertex \\
\multicolumn{1}{p{10em}}{Network optimization} & \multicolumn{1}{p{11.28em}}{Complex and large-scale networks} & \multicolumn{1}{p{9.89em}}{Complex and large-scale networks} & \multicolumn{1}{p{11.22em}}{Complex and large-scale networks} & \multicolumn{1}{p{14em}}{Complex and large-scale networks} \\
Complexity & low   & low   & comparatively low & comparatively higher \\
\hline
\end{tabular}
\label{tab:RankMethods}%
\end{sidewaystable*}

\subsubsection{Vertex Ranking}

Discovering the most important nodes in large-scale and complex networks has attained great consideration from scholars \cite{chen2012identifying,zhang2013identifying,liu2018identifying}. Recently, plenty of approaches have been designed to identify influential vertices in large-scale as well as complex networks. Some of the traditional and well-known methods are the centrality measurements \cite{gomez2019centrality}; these are degree centrality, betweenness centrality, closeness centrality, and eigenvector centrality. Additionally, PageRank, HITS, and Katz centrality are the other typical ranking methods applied in many aspects. Having considered the fact that classical methods do not perform well on big networks, scholars proposed numerous methods.

Chen \emph{et al.}~\cite{chen2013identifying} introduced a local vertex ranking approach called ClusterRank concerning the clustering coefficient of a vertex. Hu \emph{et al.}~\cite{hu2018ranking} proposed a novel method that ranks nodes to discover important ones by applying structure holes called E-Burt. As stated in \cite{hu2018ranking}, a structural hole is a gap among individuals who have no either direct or indirect repetitive relations. However, they have complementary sources of information. This method can be implemented in weighted networks. It considers three factors such as the connection strengths of the vertex locally, the number of links that connect the vertices, and the distribution of the connectivity strengths on its connecting links. To quantify the constraints of vertices while forming a structural hole, the authors in \cite{hu2018ranking} employed constraint coefficient. If a vertex has a smaller coefficient, it means the vertex can easily compose structural holes as well as it becomes the most influential. Hu \emph{et al.}~\cite{hu2018ranking} claimed that the more influential the vertices are, the stronger the disseminating capability they will have in the network.

Similarly, Wei \emph{et al.}~\cite{wei2018identifying} introduced a practical approach to identify influential vertices built upon network representation learning (NRL). NRL aims at learning disseminated vector representation for all vertices in a given network. This approach considers the structure of a given network including the overlapping communities found in the network. For this method, information distributed to several communities via vertices in community overlaps. Wei \emph{et al.}~\cite{wei2018identifying} claimed that if a vertex is a member of multiple communities compared to other vertices, then there is a high probability that this vertex will have an influence on more communities than others. According to the experiment done in \cite{wei2018identifying}, the method is pertinent to networks that are complex and large-scale.

Salavati \emph{et al.}~\cite{salavati2018bridgerank} proposed an influential node detecting method that takes into account the closeness centrality of vertices in a network. The authors proposed a ranking algorithm called BridgeRank by improving the closeness centrality measure using the local structure of vertices. The proposed method implemented as follows. First, it finds the local centrality score for each vertex. Next, it extracts one prominent vertex from each community using the centrality value. Finally, the method ranks the vertices according to the summation of the vertices' shortest path length and generate the influential vertices. According to \cite{salavati2018bridgerank}, the influential vertices have the capability of high spreading information with low computational time. Moreover, the method is suitable for complex and large-scale networks compared to other benchmark methods.

\subsubsection{Motif Ranking}

There are numerous methods with disparate approaches but similar objectives designed to quantifying similarities amongst DNA motifs. There are also approaches mainly focus on discovering, grouping, comparing, and ranking network motifs~\cite{kellis2003sequencing,gordon2005tamo}. In this section, we present some of the methods which can be applicable for BNs in their chronological order.

Having considered the lack of methods that discover motifs, match, compare, and cluster known motifs, Kankainen \emph{et al.}~\cite{kankainen2007matlign} developed a web-based tool called Matlign. Matlign fills these gaps, especially reduces repetition of similar motifs. Matlign mainly facilitates post-processing such as clustering, matching, and comparing DNA sequence motifs. Matlign is implemented on transcription factor databases which stores profiles of transcription factor binding sites. In such cases, motifs can be represented by two formats such as position frequency matrices or consensus sequences. Thus, Matlign facilitates the post-processing of discovered motifs in both formats. It also initiates from a massive amount of pre-identified motifs, and discovers, aligns, and evaluates the similarities of motifs generated by prediction tools. Consequently, the tool clusters the discovered motifs together and generates a set of non-redundant motifs. Kankainen \emph{et al.}~\cite{kankainen2007matlign} conclude that their tool outperforms other previously proposed methods based on the extensive comparative analysis they have done.

Similarly, Habib \emph{et al.}~\cite{habib2008novel} designed a method that identifies and compares discovered motifs with already-known motifs and gives a set of non-redundant motifs. The method initially adopts relevant motif discovery algorithms for detecting new motifs and filtering them in accordance with their profusion amongst the given set of sequences. Afterward, clustering and merging of newly detected motifs take place individually by considering a non-redundant group of motifs. Finally, the method ranks and identifies a non-redundant set of motifs. Having compared with other approaches, this method is more relevant to be applied in BNs.

\subsubsection{Community Ranking}

Numerous real-world BNs such as co-authorship networks, social networks, neural networks, and so on comprise community structures \cite{li2015influential}. Since the past few decades, the problem of identifying clusters/communities in a complex and large-scale network is the most crucial problem which attracts scholars' attention~\cite{fortunato2016community}. The community identification problem focuses on discovering available communities/clusters in a particular network. However, community detection approaches failed to consider the most influential communities amongst the discovered ones. Most of the approaches identify key vertices to form a community surrounding them. Identifying the top influential community plays a critical role, for instance, to find out the community which is capable of spreading information faster to other communities in a network \cite{zhan2016identification}. Moreover, \emph{Li et al.} \cite{li2017most} discussed that one vital feature of a community is the ability to propagate information for the outsiders. Another instance is that, assume that Ana is a new big data researcher and she wants to investigate some specific research problem. Hence, she wants to discover the most influential research teams from a co-authorship network in which Big Data related research issues are investigated. The discovered team supposed to be beneficial to produce quality research work. Thus, recently few research works have been done on this problem, \emph{Li et al.} \cite{li2015influential} was the first to formulate the problem of unraveling the most prominent communities in a large network. Subsequently, Doo \emph{et al.}~\cite{doo2014extracting} proposed influential community detection approach by adopting undirected network. Doo \emph{et al.} described a community's influence as ``the minimum weight of vertices in that specific community and a community with the largest influence value considered as the top influential community.'' In another work, Du \emph{et al.}~\cite{du2015tracking} proposed a community ranking method that classifies communities based on their strength, which alters over time. Moreover, Faisal \emph{et al.}~\cite{faisal2015edge} discussed remarkable scenarios that emphasized the need and significance to discover the most influential communities in a particular network. From the perspective of BNs, identifying the most influential community could reduce the complexity and computational time of the process than identifying key vertex in a whole big network.

Li \emph{et al.}~\cite{li2015influential} proposed a model ``$k$-influential community'' that can capture an influential community in a network by adopting the idea of \emph{k}-core. To begin with, Li \emph{et al.}~\cite{li2015influential} gave a formal definition of `influential' in an individual and community levels. Li \emph{et al.} suggested numerous approaches and optimization for investigating the ``finding of influential communities'' research problem. Based on their model, they introduced an online searching method aiming to unravel the ``top-$r$ $k$-influential communities'' of a given undirected network. Furthermore, for getting a fast searching process, they proposed a ``linear space index structure,'' which maintains efficient searching of the ``top-$r$ $k$-influential communities'' in an optimal time. They experimented the algorithms on different large-scale networks. Having considered the limitation (i.e., high time complexity) occurred during applying the influential community model on big networks, Li \emph{et al.}~\cite{li2017finding} proposed an improved approach called Influential Community-Preserved Structure (ICPS). ICPS reserves $k$-influential communities as well as holds linear space concerning the size of the network.

Zhan \emph{et al.}~\cite{zhan2016identification} introduced a method that discovers top-$k$ influential communities in a big network by adopting the well-known centrality measure that is Katz centrality. They considered Katz centralities to define the strength of communities. They assumed that an influential community is the one that connects to more number of communities. In such a case, information can be disseminated immediately to the largest possible number of communities available in the network. Zhan \emph{et al.} employed two main factors to rank the communities in a network. First, they compute the average katz centrality value of each individual vertex in a particular community. Second, they discover the total communities into which a particular community could propagate information. To do that, they calculate the interactions of the vertices in a community with vertices in different neighboring communities. A community with a higher value of Katz centrality is considered as the most influential community if it can able to share information to the maximum number of different communities in a network apart from the disseminator community.~\cite{zhan2016identification}.

Bi \emph{et al.}~\cite{bi2018optimal} proposed a method called LocalSearch that is an instant-optimal algorithm with a linear computational complexity. On top of that, they introduced an approach that facilitates LocalSearch in a progressive way to computing and reporting top-$k$ influential communities in a descending influence value. The subnetwork's influence value is explained as ``the minimum weight of the vertices in a subnetwork''. Unlike the method discussed previously, this does not demand to specify the value of $k$. As described in \cite{bi2018optimal}, a user has an option to end the algorithm as far as the determined influential communities have been generated.

\subsection{Partition Algorithms}

Partitioning is a decomposition technique that optimizes the handling of complex systems. Partitioning techniques decompose a big network into manageable smaller subnetworks called clusters or communities. Hence, any BN applications can be applied on the subnetworks independently to such a degree that reduces the complexity and computational costs. Partitioning methods have to minimize the linkage amongst the subnetworks.

\begin{defn} Given a network $G(V, E)$, wherein, each vertex $v \in V$, $V$ is considered as the total size of the network in terms of vertices. The problem of partitioning is to divide $V$ into $\kappa$ disconnected subnetworks $\{v_1,\ldots,v_\kappa\}$ such that it optimizes the functionality of the network, based on certain constraints.\end{defn}

While applying partition algorithms (PAs), initially the number of communities is given as an assumption as well as a network $G$ of $V$ vertices. Subsequently, PAs construct the vertices into $\kappa$ partitions ($k \leq V$), where each partition indicates a cluster/community and each vertex belongs to only one community. This shows as there is no link between clusters/communities; in essence, there is a high and low inter-community and intra-community similarity, respectively. The communities are formed on the basis of distinct partitioning measurement. The vertices within a community formed by PA have similarities amongst one another, while they have disparate relation with vertices in the other community.

Implementing partitioning algorithms on BNs is vital to address some challenging issues like detecting influential vertex from a community, recommendation, link prediction, etc. For example, identifying the most influential author from a whole big co-authorship network could be time-consuming. Thus, if we partition the network, it will reduce the computational time and complexity while discovering the influential authors.

There are some traditional partitioning methods such as CLARANS, $\kappa$-medoids, and $\kappa$-means. In the case of $\kappa$-means, each community is represented by its center. Whereas in $\kappa$-medoids, a single vertex represents a community it belongs to. We briefly discuss these methods in the following subsection.

\subsubsection*{$\kappa$-means Algorithm}

In this method, $\kappa$ is an input parameter, which is the total of communities a network $G$ assumed to have. The $\kappa$-means algorithm takes place as follows. First, it partitions vertices into $\kappa$ non-empty subnetworks, and each subnetwork represents a community/cluster. Next, $\kappa$-means computes key points as the centroid of the communities of a particular partition in which the centroid is the central point of the community. Subsequently, it assigns the remaining vertices to the community with the nearest key point as well as the center of the community. Afterward, it calculates the mean value for each community. The $\kappa$-means process works iteratively until the partitioning criterion converges \cite{han2011data}. In most cases, assuming the number of communities (i.e., $\kappa$) in advance considered as one limitation of $\kappa$-means algorithm. Moreover, as far as BNs are concerned, defining mean values for each cluster may become costly, and it makes $\kappa$-means algorithm less applicable to be implemented on BNs.

\subsubsection*{$k$-medoids Algorithm}

As the name implies, the $k$-medoids algorithm takes medoids as the most centrally placed vertex and a reference point in a community rather than a community's mean value. As stated in \cite{bhat2014k}, a medoid is ``a statistic metric which represents that data member of a data set whose average dissimilarity to all the other members of the set is minimal.'' In the $k$-medoids algorithm, non-central vertices clustered along with the most related representative vertex. PAM - Partitioning Around Medoids is a $k$-medoids algorithm that can be effectively implemented on small datasets yet failed to work well on big networks \cite{han2011data,kaufman2009finding}. The $k$-medoids algorithm are implemented as follows. The number of partitions and dataset are given. Initially, it chooses $k$ vertices as medoids. Next, it assigns non-selected vertices to their nearest medoids. Consequently, it computes the total cost of swapping vertex, which is to find a new collection of medoids. The algorithm works iteratively until no change is demanded. In this algorithm, each iteration has the computational complexity of ``$\textsc{O}(\kappa(V - \kappa)^2)$'' which makes it unfit to be applied on BNs.

However, there are extended algorithms which are proposed under the ground of $\kappa$-means and $k$-medoid algorithms. The extended approaches can be applicable in BNs. We briefly discuss the state-of-art of partitioning algorithms, which are proposed recently.

\subsubsection*{Clustering Large Application}

Clustering large application (CLARA) algorithm is considered to be an extension of $k$-medoids method. It is designed by taking into account the lack of partitioning algorithms for large datasets and with the objective to overcome the limitations of partition around medoids \cite{kaufman2009finding}.

\subsubsection*{Clustering Large Algorithm Based on Randomized Search}

Having considered the incapability of $k$-medoids method in complex and large networks, researchers proposed a method with the ground of $k$-medoids called clustering large algorithm based on randomized search (CLARANS) \cite{ng2002clarans}. CLARANS adopt the random searching technique for expediting the clustering as well as partitioning process of a large number of datasets \cite{ng2002clarans}. As mentioned earlier, CLARANS was proposed under the basis of PAM and CLARA. From the viewpoint of BNs, CLARANS is preferable as far as efficiency and effectiveness are considered.

\subsubsection*{MapReduce-based Parallel $k$-Medoids Clustering Algorithm}

Shafiq \emph{et al.}~\cite{ shafiq2016parallel} proposed a map-reduce-based clustering algorithm that can be applied on big datasets. As stated in \cite{shafiq2016parallel}, the authors considered the growing nature of real-world networks concerning velocity, volume, as well as variety. In contrast to other classical partitioning methods, this method attains parallelization despite the size of $k$-clusters which is going to be identified. As far as the experimental results found in \cite{shafiq2016parallel} considered, we believe that this method is suitable to be applied to BNs. Table \ref{tab:comparisonPA} depicts the comparison between partitioning algorithms surveyed in this paper.
\begin{sidewaystable*}
\caption{Comparison of Partitioning Algorithms. The notations $n$, $k$, and $m$ in the time complexity denote the numbers of points, clusters/medoids, and vertices in which the data is distributed in case of \cite{shafiq2016parallel}, respectively.}

\begin{tabular}{lp{10em}p{9em}p{8em}p{8em}p{9em}}
\hline
\multicolumn{1}{c}{\multirow{2}[4]{*}{\textbf{Criterion/Methods }}} & \multicolumn{5}{c}{\textbf{Partitioning Algorithms}} \\
\cline{2-6}      & \multicolumn{1}{l}{\textbf{K-means}} & \multicolumn{1}{l}{\textbf{K-medoids}} & \multicolumn{1}{l}{\textbf{CLARA}} & \multicolumn{1}{l}{\textbf{CLARANS}} & \multicolumn{1}{l}{\textbf{Reference \cite{shafiq2016parallel}}} \\
\hline
Time Complexity & \multicolumn{1}{l}{$\textsc{O}(nk)$} & $\textsc{O}(k(n-k)^2)$ & \multicolumn{1}{l}{$\textsc{O}(k(c+k)^2+k(n-k))$} & \multicolumn{1}{l}{$\textsc{O}(k^3+nk)$} & $\textsc{O}(nk/m)$ \\
Efficiency & \multicolumn{1}{l}{less } & \multicolumn{1}{l}{better than k-means} & better than the previous & better performance & comparatively more efficient  \\
\multicolumn{1}{p{7em}}{Pre-determine k} & \multicolumn{1}{l}{yes} & \multicolumn{1}{l}{no} & \multicolumn{1}{l}{no} & \multicolumn{1}{l}{no} & \multicolumn{1}{l}{no} \\
Optimization & \multicolumn{1}{l}{small networks} & \multicolumn{1}{l}{small networks} & comparatively larger networks & \multicolumn{1}{l}{large-scale networks} & BNs \\
Advantages & works well for small-scale datasets & easily understandable, the algorithm works in a fixed number of steps, less susceptible to outliers unlike k-means  & can handle larger dataset than k-means and k-medoid algorithms & gives a better result than other methods, easily handle outliers, comparatively better when implemented on  large-scale datasets & comparatively works well on BNs, scalable and effective \\
Disadvantages & predicting the k-value and comparing the quality of the clusters are challenging tasks, does not work well for BNs & high time complexity compared to k-means, not suitable for BNs & its efficiency depends on how big the network is, there is a possibility of obtaining inaccurate clusters & although it is designed for large-scale datasets, it is not as efficient & the computational time might be higher as the size of datasets increase \\
\hline
\end{tabular}%
\label{tab:comparisonPA}
\end{sidewaystable*}

\subsection{Network Embedding Algorithms}

The emerging accessibility of big networks containing billions of vertices and edges has significantly progressed network analysis. Network embedding learns an efficient low-dimensional vector representation for vertices. Due to this, big data analysts consider implementing network embedding for numerous BN applications such as community detections, link predictions, vertex clustering, recommendations, as well as network visualization. In network embedding methods, the distance amongst vertices in the vector space captures the interactions between vertices. A vertex's structural and topological features are encoded into its vector representation.~\cite{cui2018NEsurvey}.

The classical network representation commonly avails adjacency matrix, which might encompass redundant or noise information. Whereas the Network Embedding Representation Learning (NRL) tends to learn the condensed and incessant vertices' representations in a low-dimensional space. NRL not only minimizes the redundant and noisy information but also it maintains the fundamental structure information \cite{cui2018NEsurvey}. The challenges happened during network analysis such as high computation can be prevented by calculating the distance metrics on the embedding vector as well as by computing its mapping functions. Network embedding approaches overcome most of big networks representation and analysis challenges. Cui \emph{et al.}~\cite{cui2018NEsurvey} clearly illustrated the benefits of adopting network embedding over the classical approaches. In this section, we briefly explain recently proposed state-of-the-art network embedding approaches on both homogeneous and heterogeneous networks.

\subsubsection{DeepWalk and Extended Methods}

DeepWalk \cite{perozzi2014deepwalk} is a network representation learning model that uses unsupervised way to learn low-dimensional representations for vertices in social networks. In DeepWalk, graphs are supposed to be given as an input, and it provides an output of latent representations. Furthermore, DeepWalk learns representations according to the information found on the local network and it further identifies the classifications of vertices through a random walk. The principle of DeepWalk method was later extended to a semi-supervised algorithm called Node2vec \cite{grover2016node2vec}. Node2vec amends the scheme of random-walk in DeepWalk into tendentious random-walks which discovers various neighborhoods and a network structure more effectively. Node2vec is a scalable algorithm applied for nodes to learn incessant aspect representations in a network~\cite{grover2016node2vec}. Moreover, it learns the structure of vertices to a low-dimensional-featured space representation that exploits the possibilities of maintaining neighborhood of vertices in a given network.

Tu \emph{et al.}~\cite{tu2016max} designed a method having the aim to overcome the limitation of DeepWalk, which is referred to as ``Max-Margin DeepWalk (MMDW)''. MMDW overcomes the learned representation incapability of discrimination during applying to the machine learning process. MMDW is a semi-supervised NRL model that simultaneously enhances the max-margin classifier as well as the targeted social NRL. Additionally, the learned representations in case of MMDW encompass the attributes of discrimination besides the network structure. With a similar objective, another method was proposed referred to Discriminative Deep Random Walk (DDRW) \cite{li2016discriminative}.

\subsubsection{Context-Aware Network Embedding Methods}

Tu \emph{et al.}~\cite{tu2017cane} introduced a model name Context-Aware Network Embedding (CANE) assuming that a vertex could have diversified features when connecting with diverse neighborhood vertices. Thus, CANE precisely designs the semantic relationship amongst vertices. On top of that, CANE learns the context-aware embedding for each vertex, unlike other network embedding approaches proposed prior to CANE.

Ribeiro \emph{et al.}~\cite{ribeiro2017struc2vec} presented a flexible and robust framework called \textit{struc2vec} to learn the latent representation by taking into account the structural identity of vertices in a network. Structural identity is a symmetry notion in which vertices in a network are discovered based on the structure of the network and their connection to other vertices. The \textit{struc2vec} method employs a hierarchical approach to quantify vertex similarity at a distinct range. Moreover, it builds a multi-layer network for performing and generating the structural similarities as well as context for vertices, respectively.

\subsubsection{Network Embedding in Dynamic Networks}

Enormous real-world networks that are a combination of vertices and edges have a dynamic nature that changes over time. Having considered that, scholars proposed a network embedding model called Dynamic Attributed Network Embedding (DANE). DANE concerns learning a representation of the changing attributes of vertices in a dynamic network \cite{li2017attributed}. DANE is an online framework that can effectively learn representation. DANE aimed to overcome some challenges happened while embedding representation in a changing network. One of the challenges is the possibility of incomplete features of vertices and noisy correlated network that demands a vigorous learning representation. This method gives online end embedding results by using matrix perturbation theory following the consensus embedding representation. Likewise, Yang \emph{et al.}~\cite{yang2017properties} proposed a ``MultiView Correlation-learning based Deep Network Embedding'' method, shortly referred to as MVC-DNE. MVC-DNE especially contemplates the attributes of vertices as well as the overall network structure as two interconnected views in which the learned embedded representation vector returns its attributes in both views. Goyal \emph{et al.}~\cite{Goyal2018ENE} proposed a method that employs edges in the network and labels associated with the edges for learning vertex embeddings. This method considers optimizing higher-order vertex neighborhood, roles, as well as characteristics of edges re-construction error by adopting deep-architecture.

\subsubsection{Network Embedding in Heterogeneous Information Networks}

A semi-supervised approach in the heterogeneous social network helps on classification and tagging of vertices where they are of different types with their labels \cite{jacob2014learning}. In this method, different vertex types are brought together into common latent space where they share similar features. Thus, it overcomes the limitation of direct connection for understanding the correlation between vertices. Traditionally, heterogeneous networks are analyzed by mapping to homogenous, which are unable to extract the complete information. In this approach, a general assumption is that, vertices which are not directly connected are inter-dependent. These dependencies cannot be captured using a homogenous approach. Furthermore, by learning the dependencies between heterogeneous vertices, both local and global characteristics are captured.

Chen \emph{et al.}~\cite{tang2017entity} addressed the problem of calculating distance measures between the heterogeneous entities. In data-driven applications, security is dependent on the detection of anomalies. These events are heterogeneous, and most of the exiting works use heuristic techniques to find the score of the events. In \cite{tang2017entity}, the authors modeled these embedded entities into a mutual latent space based on their occurrences. Specifically, pairwise compatibility of events is observed with the use of weighted interaction of diverse entity kinds. This model makes use of ``Noise-Contrastive Estimation,'' and it works well regardless of the latent space.

Fu \emph{et al.}~\cite{fu2017hin2vec} presented a model for neural network named HIN2Vec, which is developed with the objective in representing the rich semantic information embedded in heterogeneous vertices. The proposed model accepts a set of metapaths which specify the relationships as the input. Also, it performs prediction tasks on a targeted set of relationships to learn latent vectors of vertices. This model captures a broad class of semantic relationship between nodes based on the context.

Qu \emph{et al.}~\cite{qu2018curriculum} investigated the problem of optimal order for a selection of edges in a heterogeneous star network. Heterogeneous star network comprises of a central vertex and set of attribute vertices connected to the center vertex via various types of edges. Learning vertex representation in a heterogeneous star network has a variety of applications. The other approaches did not consider the order of sampling as a critical factor. However, the optimal order plays a critical role in understanding the low-dimensional vector. Qu \emph{et al.} modeled learning node representation problem using Markov decision process along with deep reinforcement learning algorithm to capture the optimal order.

Wang \emph{et al.}~\cite{wang2018shine} proposed a signed heterogeneous information network embedding method named SHINE. Wang \emph{et al.} addressed the problem of labeling user opinion in a heterogeneous information network. Existing approaches focus mainly on the text for predicting user sentiment. Also, without explicit labels and complexity in generating labels makes the tasks of prediction challenging. Wang \emph{et al.}~\cite{wang2018shine} developed a labelled data set of user consisting of user sentiment, social relations, and profile knowledge. Then, they use signed heterogeneous information networking framework for extracting latent representation for accurate predictions. SHINE uses deep learning based embedding mechanism to understand and extract users' inclination towards the topic.

\section{Big Network Applications} \label{BNapps}

This section comprises three subsections elaborating a wide range of state-of-the-art applications of BNs, including community detection approaches in different categories, link prediction approaches as well as recommendation systems. This review provides the fellow readers with a recent image of the state of complex network field from the viewpoint of BNs.

\subsection{Community Detection}

The main target of community detection is to disclose all available communities in a network according to a specific definition of community for a given problem. A community is a collection of densely linked vertices locally and sparsely linked with global vertices. As ``community'' has been given various definition, it can be classified as follows \cite{fortunato2010community}: \begin {enumerate*} [i)] \item hierarchical clustering that unravels the multilevel community structure of a graph by discovering the likeness for each pair of nodes, \item graph partitioning that splits the nodes of a network into $k$ clusters of pre-defined threshold, \item spectral clustering that separates the graph by adopting the eigenvectors of the given graph matrix, and \item partition clustering that splits nodes into $k$ clusters in such a way that the likeness amongst nodes is maximized.\end {enumerate*}

The problem of discovering community structures of BNs is ubiquitous in diverse types of networks, for instance, biological networks \cite{sah2014exploring}. Hence, it has recently been getting attention from scholars, although it is a problem which has been studying since a longtime \cite{newman2012communities}. Discovering the community structure of a network provides vital understandings into network components, the local community impact on the global ones, influential communities, and the like. Keep this in mind, selecting a suitable algorithm to unravel the community structure of a BN can be challenging. Also, Sah \emph{et al.}~\cite{sah2014exploring} discussed that the process of discovering the accurate community structure within a network is complicated due to the inconsistent meanings of ``community'', and different outputs from different methods. As a result, most of the existing methods were evaluated on small scale networks with known number of community. Thus, after doing extensive literature review on existing community detection algorithms; we aim to recommend relatively applicable methods fitting BNs.
\begin{table*}[hbt]
  \centering
  \caption{Categories of Community Detection Methods}
\begin{tabular}{p{10.5em}p{16em}l}
\hline
\multicolumn{1}{l}{\textbf{Category}} & \multicolumn{1}{l}{\textbf{Description}} & \textbf{Algorithm} \\
\hline
Disjoint Community Detection & There is connection among communities, every node goes to one communal.  & \multicolumn{1}{p{8em}}{Infomap~\cite{fortunato2016community}, \cite{blondel2008fast}} \\
Overlapping Community Detection & There is a possibility of overlapping between communities, a node could go to numerous communities. Overlapping community detection finds some complex\newline{}structures. & \cite{masuda2016guidance}\\
\hline
\end{tabular}%
 \label{tab:CommDet}%
\end{table*}

\subsubsection{Community Detection Algorithms}

In this section, we describe relatively suitable and recently proposed state-of-the-art community detection algorithms including the traditional label propagation \cite{raghavan2007near}, fast unfolding method \cite{blondel2008fast}, and random-walk based approaches \cite{pons2006computing,de2014mixing}.

\subsubsection*{Traditional Community Detection Algorithms}

Herein, we review some traditional methods such as heuristic and label propagation community detection techniques.

\subsubsection*{Label Propagation Method}

Label Propagation Method (LPM) is designed according to label propagation, mainly focuses on detecting communities local-wise~\cite{raghavan2007near}. The algorithm begins by giving a distinctive label to each vertex and randomizes the order of vertices. LPM performs the algorithm iteratively in which each vertex embraces a label that many of its neighbors possess. The algorithm terminates as long as every vertex has a label that happens to occur more often in the network. Thus, LPM constructs a community that is a collection of vertices with akin labels \cite{raghavan2007near}.

\subsubsection*{Louvain Method}

Louvain method is a heuristic approach that initially assigns a distinct community to each vertex of a given network~\cite{blondel2008fast}. The community detection process takes place in two stages. First, the method assumes that there will be as many communities as there are vertices. And, it quantifies modularity gain by putting away a vertex from its community to other's vertex community with a positive gain. Otherwise, the vertex will not be discarded from its initial community. The algorithm repeats this process iteratively unless there is no need for improvement. Secondly, the algorithm constructs newly created network consists of the communities generated in the first stage. As stated in \cite{blondel2008fast}, the weight of links between new vertices is equal to the total summation of links' weights amongst vertices in the adjacent communities. Having done the second stage, the louvain method re-runs the first stage until no more changes of modularity are demanded. This method could be comparatively applicable to BNs as it has been previously applied to large-scale networks like phone companies.

\subsubsection*{Random-walk-based Community Detection Methods}

Among all community detection approaches, random-walk based methods inclined to discover network communities more or less accurate with the ground-truth ones \cite{abrahao2014separability}. In this section, we briefly discuss existing random-walk based community detection methods which can be comparatively applicable for BNs.

\subsubsection*{Walktrap}

This is designed with the perception that is ``random walks on a graph tend to get `trapped' into densely connected parts corresponding to communities.'' Walktrap initializes the process and mainly computes distance, consequently by analyzing the structural correlation between vertices as well as similarity amongst communities. The computed distance is used to form vertices into communities. As discussed in \cite{pons2006computing}, there will be a higher value of distance if two vertices located in different communities; otherwise, the distance will be lower. For detecting a community structure, they used a hierarchical clustering approach as well as adopted the agglomerating method. This is to reduce the computational complexity while calculating the distance. After identifying the community structure of a given network, Walktrap merges adjacent communities which have at least an edge amongst themselves.

\subsubsection*{CONCLUDE}

De \emph{et al.}~\cite{de2014mixing} proposed a random-walk-based method called CONCLUDE (COmplex Network CLUster DEtection) aiming to bring the efficiency of global methods and computational performance of local approaches together. In this method, for detecting communities, comprising the network's topological structure to heuristic algorithms is necessary. CONCLUDE introduced the concept ``$\kappa$-path edge centrality'' while performing the process of community detection. CONCLUDE does the process in two phases. Firstly, it computes the ``$\kappa$-path edge centrality'' of each edge in the graph. Thus, they proposed ``Edge Random Walk $\kappa$-path Centrality (ERW-Kpath)'' that measures the likelihood of edges by applying a random-walk with a finite length of $\kappa$. In the second phase, it computes the distances amongst the entire pairs of linked nodes in the network using the estimation value of $\kappa$-path edge centrality and assigns them as edge weights. Finally, it partitions the weighted network by adopting the Louvain Method \cite{blondel2008fast}.

\subsubsection*{Leader-based Community Detection Algorithms}

The literature on community detection shows a variety of approaches, where node centrality and graph-based methods are used widely to capture the underlying structures in the community. Realizing the basis for the community has a wide variety of applications.

Shah \emph{et al.}~\cite{shah2010community} discussed that the traditional clustering method fails to identify the precise community structures as they depend on external connectivity properties like graph-cuts. To overcome this limitation, the authors proposed a community detection approach based on leader-follower algorithm, which depends on the internal relationship of the expected community. The proposed method uses the idea of centrality in a novel fashion to differentiate leaders from followers. Further, the algorithm learns communities naturally without depending on the knowledge of the estimated number of communities.

Information networks such as protein-protein interactions in biology, call graphs in telecommunication, and co-authorship in biometrics have dense connections within the group sharing common properties while sparse connection outside the group. Likewise, khorasgani \emph{et al.} ~\cite{khorasgani2010top} proposed an algorithm that identifies all potential leaders along with their corresponding followers, i.e., communities. Eventually, communities help realize the underlying structures in social networks. Similarly, in \cite{chen2016finding}, authors proposed ``community centrality'' based on the assumption that low degree nodes surround node with a high degree. Initially using community centrality node with the highest degree (community center) is identified, later through the process of diffusion, the method generates multiple community centers with various degrees.

Yakoubi \emph{et al.}~\cite{yakoubi2014licod} introduced an efficient framework LICOD for analyzing the performance of algorithms developed for community detection. Cohen \emph{et al.}~\cite{cohen2017node} proposed a node-centric overlapping community detection algorithm (NECTAR) on the basis of the well-known local search method, i.e., Louvain method \cite{blondel2008fast}. This method is applied to overlapping community structures to deal with multi-community membership issues.

Rossetti \emph{et al.}~\cite{rossetti2017node} presented different views on node-centric approaches in an online social network both in terms of static and dynamic scenarios using algorithmic and analytical procedures. Further, with the incomplete information on network topology, node-centric or local, a community detection approach has issues in identifying the community of a given node. To overcome this, Roberto \emph{et al.}~\cite{interdonato2017node} proposed a multi-layer network-based framework by maximizing internal density to external density ratio. Meanwhile, they also proposed a biasing scheme for identification of different degrees of layer coverage diversification.

Gmati \emph{et al.}~\cite{gmati2019new} developed Fast-Bi Community Detection (FBCD) based on bipartite graphs with maximum set matching to reduce the complexity in existing algorithms. Adding on, in \cite{marquez2019overlapping}, both link and node attribute based overlapping community detection in social networks is proposed. Deng \emph{et al.}~\cite{deng2019complex} adopted Label propagation and fuzzy C-means for a community detection where initial labels are derived from neighbor nodes and revised using fuzzy C-means membership vector.

\subsection{Link Prediction}

Link Prediction (LP) estimates the presence of a link between vertices in a given network. The mechanism that dives network evolution gives a correct prediction of the network. The experiment of predicting new links is costly in biological networks such as metabolic networks or protein-protein interaction network. The experiments on real and complex networks demonstrate a different role gives an accurate prediction. The problem of link prediction is the most vital topic which is being investigated by big data mining researchers \cite{wang2015link}. LP was first introduced by Liben-Nowell and Kleinberg \cite{liben2007link} aiming to predict new future connections between vertices which could most likely appear in a network.

Moreover, link prediction is a model especially proposed for evolving networks. There is a high possibility of newly created connections as well as the deletion of existing connections in the evolving networks. For instance, in a social network like Instagram, a user may form a link whenever she/he follows or followed by a user. At the same time, they can discard links by unfollowing a user. Furthermore, link prediction plays a vital role in recommendation systems and the Internet of Things. The well-known example is a security network in which link prediction is utilized to uncover subversive communities of criminals or terrorists \cite{lu2011link}. While for human behavioral networks, link prediction is adopted to unveil and classify the movement and activities of people in the network \cite{sid2013detection}. Moreover, link prediction also has various systems replicating social connections, e.g., email networks, sensor networks, as well as communication networks.

\begin{defn} For a given network $G(V, E)$ formed at a time $t_i$, predict the further connections appeared in the network from the time the network was initially formed $t_i$ to the time the new connection created $t_n$. \end{defn}

\subsubsection{Link Prediction in A Single Layer Network}

Substantially, while implementing link prediction methods in a single-layered network, there are three classical approaches including \textit{similarity measurement methodologies}, \textit{matrix factorization methods}, as well as \textit{probabilistic graphical model approaches} \cite{cui2018survey}. In the case of similarity measurement methods, link prediction approaches predict invisible connections by computing the similarity between vertices. Hence, the two vertices with higher similarity indicate that there is a high probability of forming a future connection. There are numerous approaches proposed on the basis of the similarity measurement methodology in which there are common parameters used in the approaches. Some of the parameters are global similarity index, indices of local similarity, and quasi-local structures of a network (see Table \ref{tab:lpparam}).
\begin{table*}[htbp]
  \centering
  \caption{LP Parameters Comparison}
\begin{tabular}{p{7em}p{18em}p{11em}}
\hline
\multicolumn{1}{c}{LP Parameters} & \multicolumn{1}{c}{Functionality } & Characteristics  \\
\hline
Global index similarity & Computes similarity of vertices by making use of the global structure data & \begin{compactitem} \item High complexity \item low speed in operation  \end{compactitem} \\
Local similarity index & This taking place according to vertex's neighbors data E.g. Jaccard Coefficient in which the probability of neighbor used to compute the similarities of pairs of vertices \cite{liben2007link}. & \begin{compactitem} \item Low complexity \item low accuracy \item faster in operation \end{compactitem} \\
Quais-local structures & Considers only two vertices to do similarity measurement and the longer paths will be removed E.g, Local Path \cite{zhou2009predicting} and Superposed Random-Walk (SRW) \cite{liu2010link} & \begin{compactitem} \item Has settlements between performance and complexity \end{compactitem} \\
\hline
\end{tabular}%
\label{tab:lpparam}%
\end{table*}

Having considered vertices structural similarities and their type effect (i.e., linking behavior of vertices), a promising LP algorithm has been proposed in \cite{fan2017efficient}. The algorithm is specially designed for a heterogeneous military network in which there are different categories of vertices and edges. The authors claimed that their algorithm outperforms the other existing similarity-based methods. Because their method predicts future connections as well as it identifies pseudo connections in a given network. Gao \emph{et al.}~\cite{gao2017projection} proposed a project-based LP method specifically for a bi-partite network. Aiming to reduce the computational time complexity of LP operation, Gao \emph{et al.}~\cite{gao2017projection} came up with a new concept that is ``Candidate Node Pair (CNP)''. CNP works based on the projected graph. A projected graph is a mapping of the bi-partite network onto a uni-partite network \cite{gao2017projection}. Gao \emph{et al.}~\cite{gao2017projection} defined CNP as follows.

``Let $G=(U, V, E)$ be a bipartite graph, $B \in U$ and $x \in V$ be two vertices in $G$, and $(B, x) \in E$. Denote the U-projected graph of $G$ as $G_u =(U, E_u)$. By adding a new link $(B, x) \in U\times V$ to $G$, then construct a bipartite graph $G^{'}=(U, V, E^{'})$, where $E^{'}= E \cup {(B, x)}$. Let $G^{'}n =(U, E^{'}u)$ be the U-projected graph of $G^{'}$. If $G_u =G^{'}u$, then $(B, x)$ is a CNP in graph G by U-projection.''

While performing the link prediction, CNP is computed on the basis of the weights of patterns it contains. Furthermore, the algorithm has a linear time complexity of \textsc{O}($m$) of a bi-partite network with $n$ and $m$ vertices in two distinct parts \cite{gao2017projection}.

As mentioned earlier, there are also LP methods proposed based on \textit{probabilistic-model-oriented}. Having considered evolutionary networks, Steve \emph{et al.}~\cite{hanneke2010discrete} proposed a statistical-model-based link prediction method called temporal exponential random graph models (TERGM). Steve \emph{et al.}~\cite{hanneke2010discrete} claimed that their model performs well with promising results on dynamic networks like communication networks, gene regulation circuitry, and so on. Ji \emph{et al.}~\cite{liu2009link} proposed a link prediction model built upon two factors such as diversion delay and time attenuation in user-object based networks. Moreover, in \cite{liu2009link}, link weight is considered so that diversion delay, as well as time attenuation, will be of a great significance to forecasting invisible connections in a user-object network. Consequently, they developed ``time-weighted network (TWN)'' model by combining the factors with the lifecycle of users \cite{liu2009link}. In \cite{barbieri2014follow}, the authors presented a Bayesian-based link prediction model considering both directed and nodes-attributed network. The model has features of estimating future connections as well as it explains each estimated connection. Moreover, they proved that their stochastic model generates accurate information in predicting connections \cite{barbieri2014follow}.

The other category of LP methods is \textit{matrix factorization}. Gao \emph{et al.}~\cite{gao2011temporal} proposed a model by taking into consideration the formulation of matrix factorization. The model proposed by Gao \emph{et al.}~\cite{gao2011temporal} employs multiple information sources in time-evolving networks so as to forecast the probabilities of connections that could appear in the near future. The information exploited by the model comprises three types, including the global structure of a network, vertex's local information along with any available contents of vertices.

\subsubsection{Link Prediction in Big Networks}

Similarity measure based methods are mostly applied in complex and large-scale networks. Because learning-based LP methods such as probabilistic-based and matrix-factorization-based methods take high computational time to develop and learn training data when applying them on BNs \cite{yao2018link}. Ma \emph{et al.}~\cite{ma2017improving} analyzed and confirmed the uniqueness of the structural characteristics of different real-world networks. Having considered that, Ma \emph{et al.}~\cite{ma2017improving} proposed a link prediction method referred to an adaptive fusion model that considers various structural qualities of a network during the LP process. The model is implemented as follows. First, it defines a logical function comprising different structural features. Consequently, it employs the noted features for the adaptive determination of the weight of feature in the logistic data. Finally, it applies the determined logistic function for obtaining the invisible or missing connections in the given network. The model follows a local index in which it adopts the information of the closest as well as the next-close neighbors. The authors believed that this could reduce the computation time of their proposed algorithm. Yazdi \emph{et al.}~\cite{yazdi2018new} proposed a community structure based link prediction method with the goal of improving security-related issues that happen in social networks. The main concern of their method is to prevent inaccurate, or fraud connections recommended to user in social networks \cite{yazdi2018new}. They exploited global structure information for mapping a network into a hyperbolic environment by adopting the structure of the network community. Moreover, Louvain community detection algorithm was employed for forming vertices in distinct clusters and forecasting future connections by performing an accurate analysis of the relations of the vertices \cite{yazdi2018new}. This method can be suitable for BNs regarding link prediction as it does the process by taking into account the network's community structure. More importantly, it suggests genuine connections and controls scam recommendations. In \cite{aghabozorgi2018new}, the authors proposed a novel similarity measure based LP method where network motifs are used as a source for estimating similarity. The method is relatively appropriate to solve the LP problem of networks with billions of vertices and edges such as BN. Yao \emph{et al.}~\cite{yao2018link} presented a similarity based LP method that mainly focuses on the interaction between paths. In \cite{wang2017link}, the authors proposed a method by applying the activeness of vertices in a dynamic network. Their new active links analyze the activeness of vertices. Having taken that into account, authors in \cite{wang2017link} designed a hypothesis in which activeness of vertices and structure of the existing vertices influence the upcoming network. The activeness or popularity of edges is built upon structural perturbation method so that it differentiates active as well as in-active vertices from the network. Moreover, the perturbation method is used to unveil new connections linked with popular vertices. On top of that, their method somehow minimizes the computational time compared to other well-known link prediction approaches \cite{wang2017link}.

\subsection{Recommendation}

Recommendation system is a way of filtering information by predicting preferential products of users according to the data of their previous preferences. In essence, the recommender system tries to meet the interests and needs of the users. It is significant to manage bulky information and overcome the problem of information overloading \cite{yu2016network}. Further, it makes life easier for internet users by providing them with personalized content and appropriate services extracted from an enormous amount of information that evolves over time \cite{isinkaye2015recommendation}. With the advent of technology and emerging data, there is an increase in education resources, so a recommendation system introduced to education resource platform \cite{xia2017big}.

It is also an emerging research area that attracts much of scholars' attention, especially of computer scientists. Moreover, recommendation methods are adopted by different areas for different reasons. Recommendation methods were widely used in many application settings to suggest the services, products, and information items to consumers. For instance, they are mostly used in e-commerce for recommending products for individual users as per their preferences and/or other users history. Using a recommendation method in research collaboration networks helps to find well-experienced and productive collaborators in a certain research area one required \cite{bai2019scientific}. Recommendation methods benefit users by notifying their needs they might not have come across to; this makes recommendation methods an alternative to search algorithms. Furthermore, recommendation methods do not demand a user to enter any keywords; instead, they store users history and make use of them for a recommendation. On top of that, recommendation methods utilize link prediction techniques to facilitate the process of recommendation.

There are different approaches to design a recommendation method, such as content-based filtering, collaborative-filtering, and hybrid-filtering \cite{yu2016network}.
\begin{itemize}
  \item Content-based filtering: Recommendation methods designed on the basis of content-based filtering consider the content information to notify individual users with relatable services (e.g., products, papers, movies, songs, books, etc.) with their history of preferences. Moreover, this approach pops up suggestions by utilizing the content from entities envisioned for a recommendation. So, analysis will be made on contents such as texts, sounds, as well as images. Based on the analysis, the recommendation method built a similarity based index amongst entities as a ground for suggesting products that match with the product a target user has rated, searched, watched, visited, and bought.
  \item Collaborative-filtering: Recommendation methods designed on the basis of collaborative-based filtering notify users by collaborating information from multiple users history. Collaborative filtering based recommendation methods make way for a user to provide information about their experience on particular services and store adequate information. Later on, the provided information can be used to provide reliable recommendations to the next users. For instance, a hotel recommendation system like trip.com suggests to users as per the ratings of the hotel given by other previous customers and the target user preferences.
  \item Hybrid-filtering: Recommendation methods designed on the basis of hybrid filtering combine the features of collaborative-based and content-based filtering techniques \cite{yu2016network}.
\end{itemize}

\subsubsection{Recommendation Methods in Big Networks}

It is known that big networks, including biological networks, social networks, co-authorship networks, and the likes are composed of vertices and edges. In most of the cases, it is crucial to provide recommendations of vertices as well as edges for future connections. For instance, a collaboration network may need co-authors recommendation to form a research team on a specific research area which can be taken as a vertex recommendation problem. Recommendation methods on big networks play a vital role in the perspective of reducing time complexity. For instance, during the process of forming a research team, ranking and identifying key vertices in a whole big network, and so on. The network has turned to be pervasive modeling way in several applications such as information and social networks \cite{aggarwal2016recommender}. As a result, it is vital to understand the network structure that can be recommended depending on the circumstances in hand. In \cite{aggarwal2016recommender}, scholars discussed the varieties of scenarios that can be used during recommendation. Some of the scenarios discussed are the following. \begin {enumerate*} [i)] \item Recommendation of vertices by authority and context in which a vertex with high degree considered to be a quality one. \item Recommendation of vertices by instances in which similarity between vertices are considered. \item Recommendation of nodes by influence and content in which the vertex that disseminates information faster is more like to be recommended. \item Recommendation of links which is similar to link prediction problem.\end {enumerate*} Bear this in mind, several scholars proposed recommendation methods that can be applicable for BNs. Herewith, we discuss some selected recent and state-of-the-art recommendation methods.

Liu \emph{et al.}~\cite{liu2018context} proposed a context-aware collaborator recommendation method, intending to recommend collaborators by taking into consideration users' contextual preferences. They developed the algorithm in two modules: \begin {enumerate*} [i)] \item Collaborative Entity Embedding (CEE) network, in which researchers and research topics are characterized by vectors according to their correlation, \item Hierarchical Factorization Model (HFM), in which it discovers researchers' characteristics regarding their activeness and conservativeness.\end {enumerate*} The authors in \cite{liu2018context} claimed that these manifest researchers' strength as well as interest to work with new researchers with whom never they collaborated before. This method recommends new potential collaborators suitable for the required research topic. As they have shown in the paper \cite{liu2018context}, according to the experimental results, the method can be applicable for BNs.

Additionally, the authors in \cite{aslan2018topic} proposed a method that provides topic recommendations for authors in a bi-partite academic information network by adopting the similarity-based link prediction approach. The method estimates the likelihood of links that could appear between authors and topics in a given academic network. Yang \emph{et al.}~\cite{yang2018nearest} proposed a nearest neighbor-based random-walk algorithm that adopts the features of a random walk with restart (RWR) and PageRank. This method is designed to provide recommendations of collaborators by combining the given network features like network structure and the likelihood of walking found on the basis of the collaboration history of individuals. With the objective to enhance the performance of singular-value decomposition recommendation method, Cui \emph{et al.}~\cite{cui2018novel} presented several context-aware recommendation methods. These methods are extended according to the singular-value decomposition approach. The proposed algorithms namely referred to as context-aware-SVD (CSVD) algorithm, two-level-SVD (TLSVD) algorithm, and context-aware two-level-SVD (CTLSVD) algorithm. The algorithms perform as follows. Initially, CSVD presents ``time'' as contextual information, and filters out inappropriate recommendations. Then, the TLSVD algorithm implemented to split the rating matrix into user and item matrices. Also, it splits the user matrix as well as the item matrix into other two different matrices by employing singular-value decomposition \cite{cui2018novel}. At last, CTLSVD provides the final suitable recommendations using the combined results such as the candidate recommendations filtered using CSVD and the matrices created by using TLSVD. The authors claimed that taking ``time'' as a context improves the performance, accuracy, and effectiveness of the recommendation results CTLSVD generates at the end of the process. Having considered the fact that the tasks uploaded in crowd-sourcing systems are supposed to be completed by online workers, researchers in \cite{safran2017real} proposed a real-time recommendation algorithms that take in to account the classifications of posted tasks. This can speed up the recommendation process as well as it saves workers time they spend on selecting appropriate tasks to complete. The proposed method contains TOP-K-T and TOP-K-W algorithms. The TOP-K-T \cite{safran2017real} algorithm benefits online workers to find the top-k most appropriate tasks. The TOP-K-W \cite{safran2017real} algorithm makes the finding of the top-k most potential workers in the crowd-sourcing systems easier for the end-users. As far as the enormous amount of data and tasks take place in the crowd-sourcing system are considered, proposing a recommendation method to overcome the challenge is appreciable work. The authors believe that this work will have a valuable impact to manage crowd-sourcing systems \cite{safran2017real}.

\section{Open Issues and Challenges}\label{openchallenges}

\subsection{Dynamic Nature of Big Network}

The dynamic features of a big network are fundamental that need to be analyzed to comprehend the overall functionalities of a certain network. Moreover, the structure of networks changes depending on the dynamic nature of vertices and edges. Analyzing dynamic networks may not be as easy as managing the network properties of static-based networks. Several works like \cite{sendina2011unveiling} have been done by researchers to facilitate the investigation of the dynamic nature of networks. Those studies show as there is a significant relationship between the dynamic nature and functionalities of a particular network. Hence, it is critical to discover the network's structure that changes over time. In most of the cases, connections amongst vertices are created, removed, and re-created along with time. As an instance, in a collaboration network, connection between collaborators exists until they complete a certain task. Over time, when the task in hand is completed, the connection will be deleted. If they happened to collaborate again in the future, then the connection will re-appear. Analyzing evolutionary networks is very challenging, especially when there are billions of vertices and edges that appear/disappear over time. It is highly recommended that some tools have to be invented that make the analysis of dynamic networks easier.

\subsection{Computational Complexity}

The emerging volume of data in networks has become a very challenging task to manage from the viewpoint of space and time. The time rate to analyze big networks is not only long but also very costly and highly computational. Although various cloud platforms have been developed to store real-world big networks information, it is still an issue that should be considered. It is preferable to manage data locally, especially when the network to deal with is a dynamic one that changes its structure over time. Hence, it is crucial and wise to give special consideration to the computational complexity of algorithms mainly designed for BNs. Some scholars have attempted to propose some approaches with the objective to reduce the computational complexity of BNs. For instance, \cite{de2014mixing} and \cite{gmati2019new} proposed community detection methods by taking into account time complexity. With a similar objective, Gao \emph{et al.}~\cite{gao2017projection} and Ma \emph{et al.}~\cite{ma2017improving} proposed link prediction algorithms that could be applicable to BNs with relatively low computational time.

\subsection{Higher-order Network Blocks}

The inner structures of BNs are generally dense and complicated. With the growing scale of BNs, the basic processing unit has shifted from traditional nodes to higher-order network blocks, i.e., motifs, graphlets, subgraphs, components, etc. It has been proved that these higher-order structures are network blocks, especially in BNs. Therefore, finding more efficient ways to detect, profile, and process these higher-order network blocks is an emerging task at present. Although the higher order organization of the network has drawn scholars' attention; however, there still exist many problems to be solved.

\section{Conclusion}\label{conclusion}

The study of a complex system is getting attention in almost all disciplines from computer science to biotechnology, sociology, and so forth. On top of that, the world is ubiquitous that everything is surrounded by interrelated entities which give both large-scale and complex sets of data. These sets of data contain entities along with their connections among each other. In this paper, we introduced a new network science concept called big network. A big network comprises information vast in size with a complicated inner structure. Thus, we survey broadly in the area of big networks and give an overview of the up-to-date models, technologies, and applications of network analysis tasks concerning big networks, as well as future directions. This review paper will provide fellow researchers comprehending of the bottom line as well as critical issues on the field of network science. Moreover, it provides a guideline framework that generally contains comprehensive research topics.

\section*{References}
\bibliography{references}

\end{document}